\documentclass[twocolumn,aps,prl,superscriptaddress,showpacs]{revtex4}
\usepackage{amsmath,bm}%
\usepackage{graphicx}
\usepackage{xcolor}
\usepackage{CJK}
\newcommand{\ud}{\mathrm{d}}
\usepackage{fancyhdr}
\begin{document}

\title{Simulation of energy scan of pion interferometry in central Au + Au collisions at relativistic energies}

\author{ Z. Q. Zhang}
 \affiliation{Shanghai Institute of Applied Physics, Chinese Academy of Sciences, Shanghai 201800, China}
\affiliation{University of Chinese Academy of Sciences, Beijing 100049, China}
\author{ S. Zhang}
\affiliation {Shanghai Institute of Applied Physics, Chinese Academy of Sciences, Shanghai 201800,  China}
\author{ Y. G.  Ma\footnote{Corresponding author: ygma@sinap.ac.cn}}
\affiliation{Shanghai Institute of Applied Physics, Chinese Academy of Sciences, Shanghai 201800, China}
\date{\today}

\begin{abstract}
We present a systematic analysis of two-pion interferometry for the central Au+Au collisions at $\sqrt{S_{NN}}$ = 3, 5, 7, 11, 17, 27, 39, 62, 130 and 200 GeV/c with the help of a multiphase transport (AMPT)
model.  Emission source-size radius parameters $R_{long}$, $R_{out}$, $R_{side}$ and the chaotic parameter $\lambda$ are extracted and compared with the experimental data. Transverse momentum and azimuthal angle dependencies of the HBT radii are also discussed for central Au+Au collisions at  200 GeV/c. The results show that the HBT radii in central collisions  do not change much above   7 GeV/c. For central collisions at 200GeV/c, the radii decrease with the increasing of transverse momentum $p_{T}$ but  not sensitive to the azimuthal angle. These results provide a theoretical reference for the energy scan program of the RHIC-STAR experiment.
\end{abstract}
\maketitle 

Quantum Chromodynamics (QCD) is a basic gauge field theory to describe strong interactions. Lattice QCD calculations predicted a phase transition from hadronic matter to a deconfined, locally thermalized Quark-Gluon Plasma (QGP) state at high temperature and small net-baryon density. The Relativistic Heavy Ion Collider (RHIC) in Brookhaven National Laboratory is believed to reach sufficiently high energy densities and temperatures for the formation of QGP~\cite{L01}, and many experimental probes have demonstrated a strong-coupling QGP has been formed at RHIC for Au + Au collisions. After the system reaches to a hot-dense QGP state for short times, the system finally experiences a hadronization to form various hadrons.  In hadronic stage, the system enters chemical freeze-out stage first and finally comes to the kinetic freeze-out stage. The space-time evolution of hadronic stage can be explored by the hadron-hadron momentum correlation function, i.e. intensity interferometry technique~\cite{Rev}.
 Hanbury Brown-Twiss (HBT) interferometry of two identical pion correlations in Au + Au collisions have been measured at $\sqrt{s_{NN}}$ = 130 GeV/c~\cite{star-pion-130GeV} and 200 GeV/c~\cite{star-pion-200GeV, phenix-pion-200GeV} at RHIC. Recently energy and system size dependence of HBT correlation of pions are also reported by RHIC-STAR~\cite{star-cu-au-9-62-200GeV}. In addition, ALICE collaboration  reported on the first measurement of HBT radii for Pb + Pb collisions at $\sqrt{s_{NN}}$ = 2.76 TeV/c at Large Hadron Collider (LHC)~\cite{alice-pion-2.76TeV}.
These experimental results invoke some theoretical works. Hydrodynamic model can describe reasonably well the momentum-space structure of the emitting source and elliptic flow~\cite{hydro-flow-hbt}. However, the source information from HBT correlations which was extracted by hydrodynamic model~\cite{hydro-flow-hbt, hydro3D-hbt} failed to describe experimental results, i.e.  called "HBT puzzle". AMPT model  can also present the HBT correlations for  two pions~\cite{AMPT-HBT-pion} and for  two kaons~\cite{AMPT-HBT-kaon}. The "HBT puzzle" was also discussed in the ultrarelativistic quantum molecular dynamics (UrQMD) model~\cite{UrQMD-HBT-puzzle} and by this model the two pion HBT correlation in Pb + Pb collisions at $\sqrt{s_{NN}}$ = 2.76 TeV was reported in Ref.~\cite{UrQMD-HBT-LHC}.

On the other hand, the Beam Energy Scan (BES) program at RHIC was launched with the specific aim to explore
the QCD phase diagram. Particular emphasis was given to the search for phase boundary and the location of the critical point \cite{Ody}. The QCD phase diagram can be accessed by varying temperature $T$ and baryonic chemical
potential $\mu_B$, which can be achieved by varying the colliding beam energy experimentally. Motivated by the importance of the hadron-hadron momentum correlation technique as well as BES program,  here we  present a multiphase model ~\cite{L02}  simulation  for the pion interferometry  in  central Au + Au collisions at different relativistic energies in this paper.

The AMPT model includes four main components: the initial conditions, scatterings among partons, conversion from the partonic to the hadronic matter, and hadronic interactions. The initial conditions, which contain the spatial and momentum distributions of minijet partons and soft string excitations, are obtained from the HIJING model~\cite{L03,L04,L05,L06}. Partonic interactions are modeled by ZPC~\cite{L07}, which at present includes only two-body scatterings. The AMPT model has two versions. One is the default AMPT model~\cite{L08,L09,L10,L11,L12,L13,AMPT-HBT-pion}, where partons are recombined with their parent strings when they stop interacting, and the resulting strings are converted to hadrons using the Lund string fragmentation model~\cite{AMPT-HBT-kaon,L16,L17}. The other is the AMPT model with string melting~\cite{L18,L19,L20}, where a quark coalescence model is used instead to combine partons into hadrons. The dynamics of the subsequent hadronic matter is described by the A Relativistic Transport (ART) model~\cite{L11,L12} and extended to include additional reaction channels which are very important at high energies. In the present work we use the AMPT version with the string melting scenario.

Hanbury Brown-Twiss  interferometry of two identical pions can directly access the space-time structure of the emitting source formed in heavy-ion collisions, providing a  probe of the system evolution dynamics. To measure multi-dimension source sizes, relative momentum ($k$) is decomposed into standard side-out-long axis~\cite{L21}: where $k_{long} (k_l)$ represents for a component parallel to the beam-axis, $k_{out} (k_o)$ for the one parallel to the transverse momentum of the pair $(\mathbf{K}_{T}=(\mathbf{p}+\mathbf{q})/2)$, and $k_{side} (k_s)$ for the one orthogonal to both $k_{long}$ and $k_{out}$~\cite{L22}. 
The ``HBT puzzle" from hydrodynamical models might arise because the combination of several effects: mainly prethermalized acceleration, using a stiffer equation of state, and adding viscosity as claimed in Ref.~\cite{L24}. 
Knowledge or reasonable assumption of duration and freeze-out shape of source can improve to understand the "HBT puzzle", which need more works in both experiments and theories.

The two-boson correlation function is given by
\begin{widetext}
\begin{eqnarray}
 C_{2}(\mathbf{p},\mathbf{q})-1 &=& \frac{\int \ud ^{4}xS(x,\mathbf{K}) \int \ud ^{4}yS(y,\mathbf{K})\text{exp}(2ik \cdot(x-y))}{\int \ud^{4}xS(x,\mathbf{p}) \int \ud ^{4}xS(y,\mathbf{q})}  \\
   &\approx&  \frac{\int \ud ^{4}xS(x,\mathbf{K})\int \ud ^{4}yS(y,\mathbf{K})\text{exp}(2ik \cdot(x-y))}{\lvert \int \ud ^{4}xS(x,\mathbf{K}) \rvert ^{2}}
   \label{fun_C2pq}
\end{eqnarray}
\end{widetext}
where $\mathbf{K}=(\mathbf{p}+\mathbf{q})/2$ and $\mathbf{k}=(\mathbf{p}-\mathbf{q})/2$. 
We calculate the correlation parameters by performing a $\chi^{2}$ fit of the three-dimensional correlation function $C_{2}(k_{o},k_{s},k_{l})$ to a Gaussian~\cite{L26} as,

\begin{equation}
C_{2}(k_{o},k_{s},k_{l})=1+\lambda \text{exp}(-k^{2}_{o}R^{2}_{o}-k^{2}_{s}R^{2}_{s}-k^{2}_{l}R^{2}_{l}).
\label{fun_C2k}
\end{equation}
Here, $\lambda$ is  often referred to as an incoherence factor~\cite{Rev}. The parameter $\lambda$ can represent the correlation strength. Theoretically it can be less than unity due to partial coherence of strong interaction, long-lived resonance decays and the non-Gaussian form of the correlation function~\cite{Rev,star-pion-200GeV}.

The correlation functions are calculated from the phase space distributions of pions at freeze-out using the CRAB (the CoRrelation After Burner) \cite{Pratt}. Given a model for a chaotic source described by $S(x,\mathbf{K})$, such as the transport model described above, Eq.~\ref{fun_C2k}~\cite{L27} can be employed to compute the correlation function.

\begin{figure}
\includegraphics[width=1.0\columnwidth]{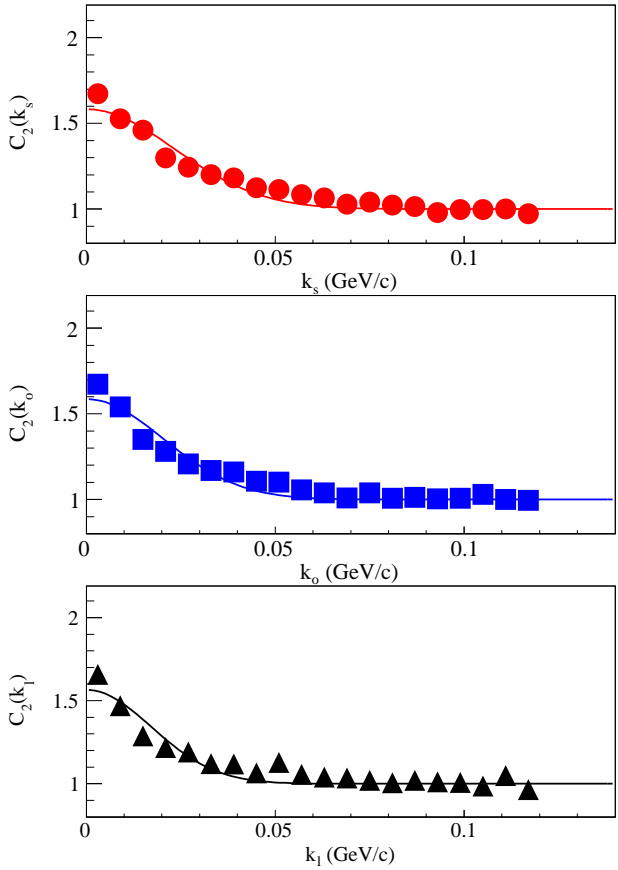}
\caption{\footnotesize  (Color online) The two pions HBT correlation function for 0-10\% centrality Au + Au collisions at 200 GeV/c in the AMPT model fitted by Eq.~\ref{fun_C2k}.}
\label{FigC2k}
\end{figure}

Here we present results of a systematic study of two-pion interferometry in central Au+Au collisions. As pion is one of the main production particles in relativistic nucleus-nucleus collisions, we use $\pi^{+}$ and $\pi^{+}$ as correlation particles. In this calculation, we use the similar event selection method in experiment~\cite{star-pion-200GeV} for centrality cuts, where the centrality was characterized according to the multiplicity of charged hadrons. In transport model, Glauber model~\cite{glauber-model} is always  employed to calculate the number of participants to define centrality, which is discussed in detail in our previous work~\cite{song-sys-size-dep}. The kinetic variable of pseudorapidity ($\eta = \frac{1}{2}\text{ln}\left(\frac{|\mathbf{p}|+p_{z}}{|\mathbf{p}|-p_{z}}\right)$, $\mathbf{p}=(p_x,p_y,p_z)$) is limited to (-1,1) for investigating mid-pseudorapidity physics. With no special statement, we selected the collision centrality of 0 to 10\% and pseudorapidity region of $\lvert\eta\rvert<1$. Figure~\ref{FigC2k} shows the two pions ($\pi^{+}+\pi^{+}$) HBT correlation functions in three-dimension for 0-10\% centrality Au + Au collisions at 200 GeV/c in the AMPT model with the help of CRAB. And the correlation functions are fitted by Eq.~\ref{fun_C2k} for the three-dimension. It  presents nice quality of the fitting in the algorithm of CRAB and Eq.~\ref{fun_C2k} based on a Gaussian ansatz.

\begin{figure}
\includegraphics[width=0.95\columnwidth]{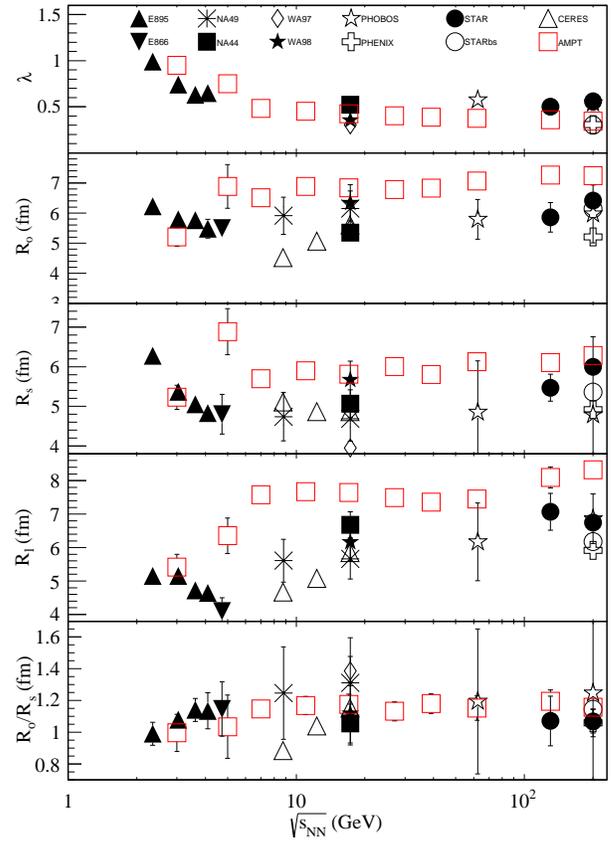}
\vspace{-0.1truein}
 \caption{\footnotesize (Color online) Energy dependence of HBT parameters for central Au+Au, Pb+Pb, and Pb+Au collisions at mid-pseudorapidity and  $<k_{T}>\approx0.2$GeV/c form AGS, SPS, RHIC and AMPT model \cite{star-pion-130GeV,star-pion-200GeV,L31,L32,L33,L34,L35,L36,L37,L38}. Error bars on NA44, NA49, PHOBOS, CERES and STAR results at $\sqrt{S_{NN}}$ = 130 GeV/c and 200 GeV/c include statistical and systematic uncertainties; error bars on other results are statistical only.}
\label{FigHBTRadii_Lambda_sNN}
\end{figure}

\begin{figure}
\includegraphics[width=1.05\columnwidth]{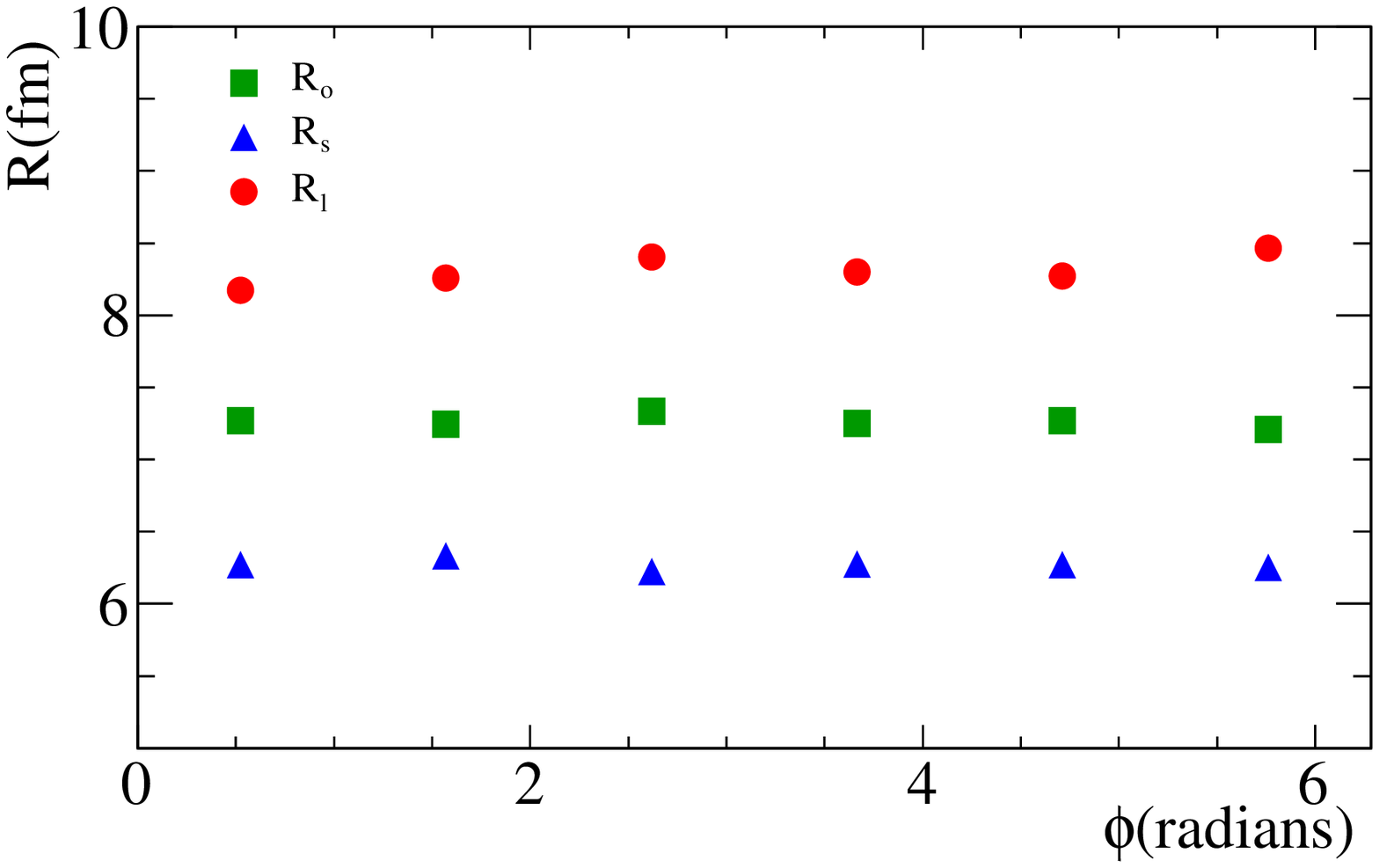}
\vspace{-0.1truein} \caption{\footnotesize  (Color online) The HBT radii relative to the reaction plane angle for 0-10\% centrality for AMPT model at 200GeV/c.}
\label{Fig_phiDep}
\end{figure}

Figure~\ref{FigHBTRadii_Lambda_sNN} shows the HBT radii, the chaotic parameter $\lambda$ and the ratio $R_{o}$/$R_{s}$ from the AMPT model and their comparisons with the experimental results. The chaotic parameter  and the ratio  from the AMPT model are consistent with the experimental results~\cite{star-pion-130GeV,star-pion-200GeV,L31,L32,L33,L34,L35,L36,L37,L38}, but the HBT radii from the AMPT model seems larger than those from the experimental results. Moreover, the radii do not show much dependence on the beam energy from 7 GeV to 200 GeV.

\begin{figure}
\includegraphics[width=1.0\columnwidth]{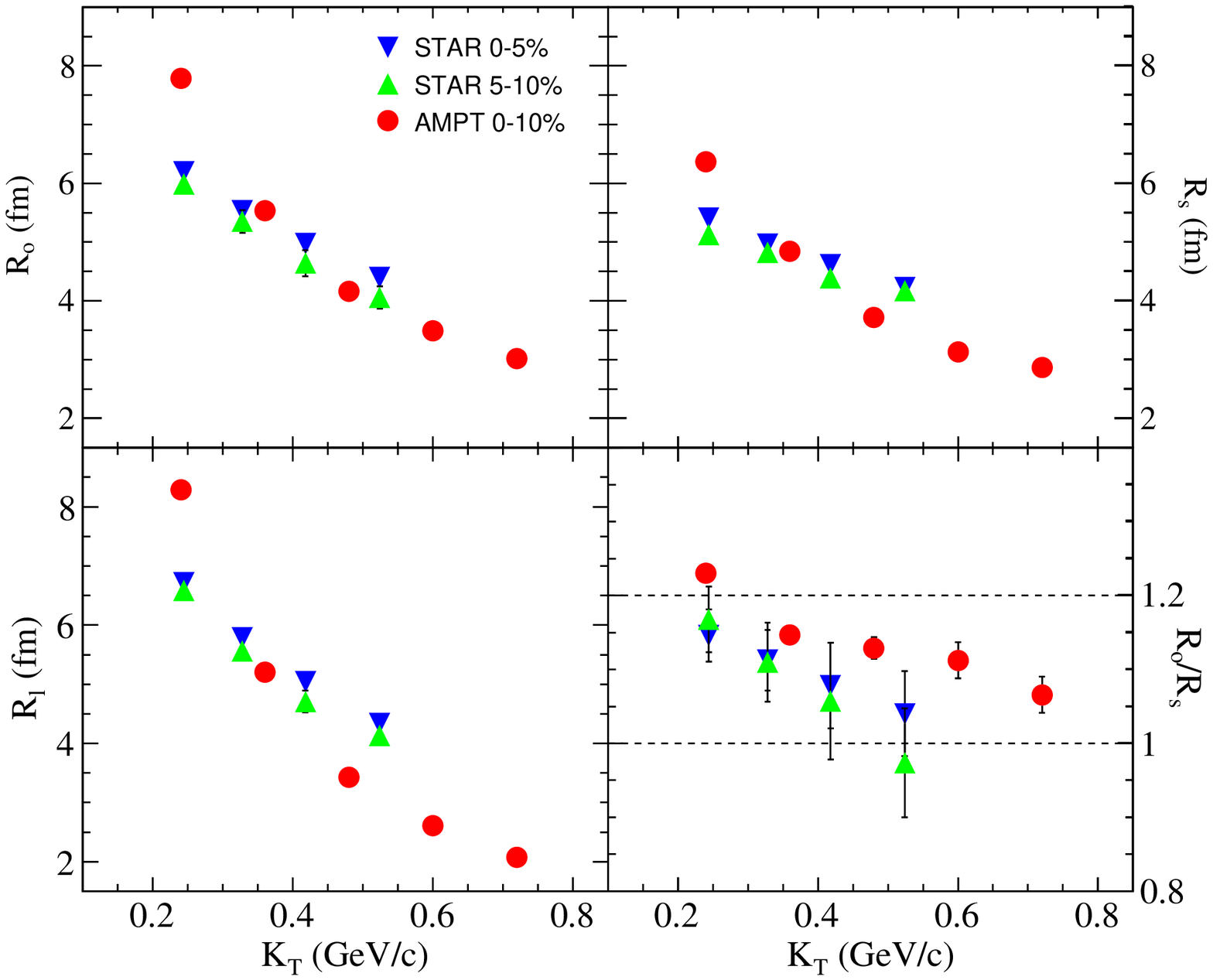}
\vspace{-0.1truein} \caption{\footnotesize  (Color online) The $K_{T}$ dependence of the HBT radii for 0-10\% centrality by the AMPT model, comparing with the STAR data at 0-5\% and 5\%-10\% centrality  at 200GeV/c.}
\label{Fig_KTDep}
\end{figure}

Figure~\ref{Fig_phiDep} shows azimuthal angle dependence of HBT radii in 0-10\% centrality at 200 GeV/c with the AMPT model. $R_{o}$, $R_{s}$ and  $R_{l}$ are not so sensitive to the azimuthal angle $\phi$. Considering that we only select 0-10\% centrality, it is reasonable not to see a strong dependence on $\phi$. These results are consistent with those from STAR for the same system~\cite{star-pion-200GeV}.

Figure~\ref{Fig_KTDep} shows the $K_{T}$ dependence of the HBT radii for 0-10\% centrality by the AMPT model, and 0-5\%, 5\%-10\% centrality for STAR results at $\sqrt{S_{NN}}$=200 GeV. It can be seen from Figure~\ref{Fig_KTDep} that the HBT radii and the ratio  from AMPT model decrease with the increasing of transverse momentum  and the  ratio $R_{o}/R_{s}$ is around 1.0-1.2. Qualitatively speaking, high transverse momentum mesons are ejected from the emission source earlier, while the low transverse momentum meson emits  lately, therefore we can see the expansion of the emission source by the $K_T$ dependence of HBT radii.

In conclusion, we calculated $\pi^{+}+\pi^{+}$ correlation function and extract the emission radius parameters for central Au+Au collisions in wide  RHIC energies. It shows that the chaotic parameter  and the ratio $R_{o}$/$R_{s}$  from the AMPT model are consistent with the experimental data, but the HBT radii from the AMPT model are larger than experimental results. We also present an analysis of the transverse momentum and azimuthal angle dependencies of the HBT radii in central Au+Au collisions at  200 GeV/c. The results show that HBT radii are not sensitive to the azimuthal angle and it decreases with the increasing of transverse momentum p$_{T}$ in central collisions.

We are grateful to Dr J. H. Chen, P. Zhou, L. X. Han and G. Q. Zhang for many valuable discussions.
This work was supported in part by the 973 program under contract 2014CB845401,  the National Natural Science Foundation of
China under contract Nos. 11035009, 11220101005, 10979074,  11105207, 11175232
and the Knowledge Innovation Project of
the Chinese Academy of Sciences under Grant No. KJCX2-EW-N01.


\end{document}